\documentstyle[12pt,a41,rotating]{article}

\newcommand{\be}{\begin{equation}}
\newcommand{\ee}{\end{equation}}
\newcommand{\bd}{\begin{displaymath}}
\newcommand{\ed}{\end{displaymath}}
\newcommand{\baa}{\begin{array}{lll}}
\newcommand{\eaa}{\end{array}}
\newcommand{\ba}{\begin{eqnarray}}
\newcommand{\ea}{\end{eqnarray}}

\begin{document}

\noindent
CPT-97/P.3564   \\
DESY 97-232 \\
hep-ph/9712207\\
November 1997 \\
\vspace{3cm}

\begin{center}
{\LARGE\bf Theory Meets Experiment    \\
for the Determination of $\Delta G/G$
\footnote{Summary of the Discussion Session at
 the Topical Workshop on 'Deep Inelastic
  Scattering off Polarized Targets', Zeuthen, Sept. 1-5, 1997} }

\vspace{1.5cm}
{D. von Harrach$^a$,  W.-D. Nowak$^b$  and  J. Soffer$^c$}

\vspace*{1cm}
{\it $^a$Institut f\"ur Kernphysik Universit\"at Mainz,\\
Staudingerweg 7, D-55099 Mainz, Germany}\\

\vspace*{5mm}
{\it $^b$DESY -IfH  Zeuthen, \\
Platanenallee 6, D-15738 Zeuthen, Germany}\\

\vspace*{5mm}
{\it $^c$Centre de Physique Th\'eorique - CNRS - Luminy,\\
Case 907 F-13288 Marseille Cedex 9 - France}\\

\vfill

\end{center}

\begin{abstract}
\noindent
We briefly summarize the main points made during the
discussion session on the determination of $\Delta G/G$, where all
Workshop participants were strongly encouraged to contribute. 

\end{abstract}

\vspace*{2cm}

\newpage

\section{Introduction}

\vspace{1mm}
\noindent
Several challenging questions remain of central importance in the
field of spin physics, and clearly one of them concerns the spin
structure of the nucleon. In 
spite of the remarkable progress accomplished over the last ten
years or so, both on the experimental side and on the theoretical
side, we are still lacking a very precise answer to the fundamental
question: 
\begin{center}
{\it What are the different components of the nucleon spin?} 
\end{center}
The recent measurements at CERN \cite{RM}, DESY \cite{DH}
and SLAC \cite{CY} of polarized Deep
Inelastic Scattering with various polarized targets have led to the
extraction of the spin-dependent nucleon structure functions
$g_1^{p,n}(x,Q^2)$, for proton and neutron, in a limited $x,Q^2$
kinematic range. This information allows the determination of the
various polarized parton distributions, which can be done with a
reasonable accuracy only for valence quarks $u$ and $d$, namely
$\Delta u_v$ and $\Delta d_v$. However, the situation remains largely
ambiguous and controversial for sea quarks and gluons. This is the
reason why several experiments are actually planned or under
discussion to measure the gluon polarization $\Delta G$ in view of 
obtaining reliable information on this important physical
quantity. \\
Details on these experiments can be found in the below two-fold table.
This table recalls for each experiment a fair number of
characteristic features, both theoretical and experimental. The upper
part of the table contains two approved experiments to start running
by the year 
2000 or so (RHIC \cite{GB}, COMPASS \cite {KM}); in addition two
recently introduced projects are summarized (E156 at SLAC \cite{PB}
and APOLLON at DESY \cite{MD}) \footnote{We note that, as of Oct. 1997,
both E156 and APOLLON were not recommended for immediate realization by
the respective physics committees.}.
The lower part of the table refers to experiments under discussion
for the more distant option of a possible polarized HERA collider
(collider option \cite{CO}, fixed target option \cite{FT})
which are bound to the uncertainties associated to what strategy will 
be eventually chosen for the future physics
programme at HERA in about five years from now.

The purpose of this discussion session was, on one hand, to
encourage theorists to review and clarify the theoretical basis for
each experimental method and to argue in favour of, or against it. On
the other hand, one was also hoping to hear about new arguments which
could cast (or remove) serious doubts on (from) any straightforward or naive
interpretation of some of these future experimental results. 
The discussion was organized by suggesting essentially three broad
provocative questions which were partly answered, but which also
generated  new challenging queries as we will see now.

\section{What Does Theory Make Out of a Measured \\
Polarized Cross Section ?}

\vspace{1mm}
\noindent

This question came about, because some people were concerned by the
fact that, perhaps, very different experiments may not measure the
same physical quantity $\Delta G$. Here we are touching the general
question of factorization/regularization scheme dependence, which also
applies in the spin average case, to the determination of the
unpolarized gluon distribution $G$ and, more generally, of any parton
density. From the accurate measurement of the structure function $F_2$
at HERA, one can extract indirectly the gluon density $G$, via the
$Q^2$ evolution, which must be identical to the one obtained more
directly, for instance, in the analysis of the prompt photon
production data, in $pp$ and/or $p\bar p$ collisions. Of course, for a
unique interpretation, this implies that in a phenomological analysis,
in particular at the next-to-leading order, the coefficient functions
and the parton densities are defined in a consistent scheme. One is
usually choosing the modified minimal substraction scheme, the
so-called  $\overline{{\mbox{\rm MS}}}$ scheme. From this point of
view $\Delta G$ is exactly like any other parton density. However, it
is important to stress that there is a major difference between
$\Delta G$ and $\Delta q$. In the Operator Product Expansion, whereas
the first moment of the quark polarization $\Delta q$ is associated to
a twist-two gauge-invariant local operator, this is not the case for
the first moment of $\Delta G$. This important theoretical subtlety
does not mean that $\Delta G$ is not a measurable physical quantity. 

In addition to the quark and gluon helicity components of the nucleon
spin, it is also legitimate to think about possible contributions from
the quark and  gluon orbital angular momentum $L_q$ and $L_G$,
respectively. One can write down the following simple spin sum rule 
\begin{eqnarray}
\label{ssr}
{\textstyle\frac{1}{2}} ={\textstyle\frac{1}{2}} \Delta \Sigma + \Delta G + L_q  +  L_G ~,
\end{eqnarray}
where $\Delta \Sigma$ is the sum of the quark (antiquark) helicity
components. It can be shown \cite{Ji1} that asymptotically the quark
and gluon contributions, namely $J_q=\frac{1}{2} \Delta \Sigma + L_q$
and $J_G= \Delta G +L_G$, are gauge-invariant and in the ratio
3$n_f$/16, where $n_f$ is the number of active fermion
flavors. However, this does hold at finite $Q^2$, so the knowledge of
$\Delta \Sigma$ and $\Delta G$ at today or tomorrow accessible
kinematics is not sufficient to speculate about
the importance of $L_q$ and $L_G$; there are suggestions on how to
measure them, {\it e.g.} by means of Deeply Virtual Compton Scattering
\cite{Ji2}. 

\section{What is the Cleanest Way to Measure $\Delta G/G$ ?}

\vspace{1mm}
\noindent

 Here we have heard comments on more general theoretical problems, in
 particular, on some tricky points concerning the models corresponding
 to the various processes indicated in the Table, which are usually
 ignored to make life simpler. In the case of APOLLON, the physical
 process is exclusive diffractive $J/ \psi$ leptoproduction with a
 polarized target. It is claimed to be a promising way to get directly
 $\Delta G /G$, according to a two-gluon exchange
 model \cite{R}. Clearly, the first objection to be made to this simple
 result is the fact that it is a leading-order approximation and it
 might be spoiled by higher order corrections.They seem to be very
 hard to calculate and, in addition to several reservations which were
 mentioned, the discussion led to the conclusion that inclusive open
 charm, as proposed by COMPASS, is certainly more reliable and
 therefore more desirable. We were told that, as a general rule, "the
 more exclusive is a process, the less theorists understand it !". On
 the other hand, one can argue, and this was done, that if exclusive
 diffractive $J/ \psi$ leptoproduction works in the unpolarized case,
 this is a good enough reason to expect it to work in the polarized
 case. Concerning COMPASS, we can ask whether there are hidden
 theoretical problems. One of them is soft gluon radiation, which is
 not yet solved but is on the way and should be easier since we are
 dealing with almost real photons. Large corrections might be found,
 like for example, in the case of heavy quark pair production in
 polarized photon-photon collisions \cite{JT}. Perhaps one should also
 worry about collinearity and the fact that a reliable QCD prediction
 requires each of the heavy quarks to be produced with a significant
 scattering angle, but this did not appear to be a serious problem. 

If we now turn to RHIC, the golden channel is inclusive prompt photon
production. The question of the isolation cut is an important one, but it
is an experimental problem. Remember that the interpretation of this
process at high energy is not as clear as it might appear. For example
in the case of the cross section from the recent FNAL data by E706, a
standard NLO QCD description fails and one needs to take into account
some sizeable intrinsic $k_\perp$ effects. We hope that this unpleasant
feature will drop out in spin asymmetries, which are cross section
ratios. One can also ask how much we can trust the extrapolation to
RHIC energy, of the present determination of the quark polarizations,
which enter in the extraction of $\Delta G/G$. The answer to this is
that RHIC will also measure $\Delta q/q$ independently.  

\section{What if $\Delta G=0$ ?}

\vspace{1mm}
\noindent

The answers to this speculative question reflected several attitudes,
which were expressed in the discussion. First, it was claimed that
this result would be a triumph for chiral bag models and here one can
recall, in particular, the Skyrme model in which the nucleon
corresponds to a soliton solution of a chiral Lagrangian \cite{BEK},
where one expects $\Delta G=\Delta \Sigma=0$ and therefore
$L_{q+G}=1/2$. However, present data seems to exclude  $\Delta
\Sigma=0$ , but the need for a measurement of the orbital angular
momentum component was stressed once more. Clearly $\Delta G=0$ would
not be a catastrophy and one should remember that present data is not
inconsistent with the so-called "Chihuahua $\Delta G$"\cite{ER}. A
small $\Delta G$ is certainly less unpleasant than a very large one,
which would create a real embarrassment. Finally, one should not
forget the importance of the $Q^2$ evolution since, if one finds $\Delta
G=0$ at two different $Q^2$ values, we would have to throw away
perturbative QCD. \\

\subsection*{Acknowledgements} 
We thank A.~Br\"ull, G.~Bunce, S.~Forte, R.L.~Jaffe, E.~Leader, J.~Lichtenstadt,
G.~Mallot, P.~Mulders, W.~van~Neerven, P.~Ratcliffe, E.~Reya,
A.~de~Roeck, D.~Sivers, O.~Teryaev, A.~Tkabladze, W.~Vogelsang, and A.~Vogt
for their participation in the lively discussion. We are greatful to A.~Deshpande
and V.~Hughes for their kind permission to further develop their table of
experiments aiming at measuring $\Delta G/G$ in the future, which was originally
developed for a similar discussion session during the 1996
Amsterdam Spin Conference \cite{VWH}. The support of
P.~Bosted, G.~Bunce, M.~D\"uren, G.~Mallot, and G.~R\"adel in providing
us with the most actual numbers to complete the current version of this
table is highly acknowledged.

\newpage

\hspace*{-0.8cm}
\begin{turn}{90}
\hfil
\begin{tabular}{||c||c|c|c|c||}
\hline\hline
\multicolumn{5}{||c||}{~}\\
\multicolumn{5}{||c||}{\Large Experiments Planned To Measure $\Delta G$}\\
\multicolumn{5}{||c||}{~}\\
\hline\hline
	   &	    &	& & \\
\bf{EXPERIMENT}&\bf{SLAC-E156}& \bf{COMPASS @CERN} &\bf{RHIC} & \bf{APOLLON}\\
& & &	&	 \\
\hline
\bf{Quantity} & 
${\cal A}^{c\bar{c}} (also {\cal  A}^{J/\psi})$& 
${\cal  A}_{\vec{\mu}\vec{N}}^{\mu c\bar{c}}$	&
 ${\cal A}_{\vec{p}\vec{p}}^{\gamma jet}$ & 
${\cal A}_{\vec{\gamma}\vec{N}}^{J/\psi}$\\
\bf{Measured}			
& $10~ \gamma$ energies					
& up to$~4~\nu~$bins			
& several $x_G~$bins
&				\\
\hline
& 
${\gamma}+{N}\rightarrow c\bar{c}$& 
$\vec{\mu}+\vec{N}\rightarrow \vec{\mu}+c\bar{c}$&
&
$\vec{\gamma}+ \vec{N} \rightarrow c\bar{c} \rightarrow J/\psi$\\
\bf{Processes} & 
$c\rightarrow \mu~(24\%), high ~p_{T}$ & 
$c\rightarrow D^0 \rightarrow K^-\pi^+$~($4\%~ BR$)& 
$\vec{p}+\vec{p}\rightarrow\gamma+$jet &
$J/ \psi \rightarrow \mu^+ \mu^- ~(BR~ 6\%)$\\
& 
$ (also ~ J /\psi \rightarrow \mu^+ \mu^- )$&
 Also $D^{*+} \rightarrow\pi^+D^0$ & 
&\\		    
\hline
\bf{Kinematical} & 
Coherent $\gamma's,$~Q$^2=0$ &
 Quasi-real $\gamma's$ Q$^2 \approx 0$ &
&
 Compton $\gamma's$ ,$~Q^2=0$\\
\bf{range}& 
$E_{\gamma} \sim$ 16, 19, 22, ..., 45 GeV& 
$35<\nu<85$&
&
$E_{\gamma} \le $ 18 GeV\\	
& 
 $0.1<x_G<0.4$ & 
$0.06<x_G<0.35$& 
$0.03<x_G<0.4$&
0.3 $<x_G<0.5$ 	\\
\hline
\bf{Theoretical} & 
\multicolumn{3}{c|}{~~~~~~~~~~LO available, NLO in
  progress~~~~~~~~~~~~~~~~~~~~~~~~~~~~~~~~~~~~~~~For $qg
\rightarrow \gamma (qjet)$}&
CSM		\\
\bf{Basis} \&			& For $\vec{\gamma}+\vec{N}\rightarrow
c\bar{c}$		& For  $\vec{\mu}+\vec{N}\rightarrow
\vec{\mu}+c\bar{c}$		& BG from $q\bar{q}
\rightarrow \gamma (gjet);$	&
LO available, \\
\bf{Uncertainties}		& \multicolumn{2}{c|}{$c$ quark mass
  uncertainty}
& Should know $\Delta q$	&NLO to be done	\\
\hline
\bf{Kinematical}& 
$p_T^\mu \ge 0.5$ GeV & 
Events at $D^0$ mass & 
$5<p_T<30$&
Cut: z $<$ 0.9 (inelastic)  \\
\bf{Constraints} & 
$p_\mu > 3$ GeV&
&
&
z $>$ 0.9 (elastic)	 		\\
\hline
\bf{Experimental} & 
Backgrounds from $\pi$,K& 
Combinatorial 						
& Identify direct $\gamma's$;&
 		\\
\bf{Problems}& 
decays, Bethe-Heitler& 
background from $K/\pi$	& 
Contamination from&
 Laser Cavity\\
&								
& 
$B/S \approx 4~$& 
$\pi^o \rightarrow \gamma\gamma$&	\\
\hline
\bf{Statistical Error}&
0.009, 0.006, 0.008,&							
&
& 					\\
\bf{on} ${\cal A}$ & 
... 0.030&
$\delta {\cal A}_{\gamma}^{\mu c\bar{c}}=0.05$ for full data & 
$\delta {\cal A}=0.002-0.04$&
$\delta {\cal A}^{J/\psi}=0.05$		\\
\hline
\bf{on} $\Delta G/G$ & 
0.01, 0.02, ..., 0.06 & 
$\delta <\Delta G/G>= 0.10$ & 
$\delta(\Delta G/G)= 0.01-0.3$&
$\delta(\Delta G/G)= 0.15$	\\
\hline
\bf{Systematics} & 
Beam, target polarizations,&
 Beam \& target	& 
Beam polarization $\pm 6\%$&
Analysing Power		\\
& 
dilution (total $\sim$ 6\%)& 
polarizations	 $\pm 4\%$ & 
False asymmetries small	&
($J/\psi \rightarrow \Delta G)$	\\
\hline
\bf{Status}& 
Decision Sept 12,1997& 
Approved by SPSLC & 
RHIC complete with&
Draft Proposal \\
 &								&
 
 & 
 Siberian snakes in 1999&  \\
\hline
\bf{Time scale}& 
1999 - 2000 & 
$\geq Year ~2000$ & 
Accelerator and detectors&
2000		\\		 
& 								
& 								
& 
ready after year$~2000$&
		\\
\hline
\bf{Remarks} & 
Data taking : & 
Apparatus shared &
 Apparatus shared&
Error for 1 year 			\\
& 
4 months & 
with hadron program & 
with heavy ion program	&
running	\\
\hline\hline
\end{tabular}
\hfil
\end{turn}

\clearpage

\newpage
\hspace{-0.5cm}
%
\begin{turn}{90}
\hfil
\begin{tabular}{||c||c|c|c||}
\hline\hline
\multicolumn{4}{||c||}{~}\\
\multicolumn{4}{||c||}{\Large Experiments Planned To Measure $\Delta G$}\\
\multicolumn{4}{||c||}{~}\\
\hline\hline
								& \multicolumn{2}{|c|}{~}					&										\\
\bf{EXPERIMENT}							& \multicolumn{2}{|c|}{{\bf POLARIZED HERA}}			& {\bf HERA-N}									\\ 
								& {\bf Inclusive}						& {\bf Exclusive (2-jets)}				&			\\
\hline
\bf{Quantity}							& $g_1^p(x)$					& ${\cal A}_{\vec{e}\vec{p}}^{e(2~jets)}$	&  ${\cal A}_{\vec{p}\vec{N}}^{\gamma jet}$ \& ${\cal A}_{\vec{p}\vec{N}}^{J/\psi jet}$	\\
\bf{Measured}							& wide $x$-$Q^2~$range				& several $x_G~$bins 	& several $x_G~$bins	\\ 
\hline
\bf{Process}							& Polarized inclusive						& $\vec{e}+\vec{p} \rightarrow 2~$jets	& $\vec{p}+\vec{N}\rightarrow\gamma(J/\psi)+jet$ 	\\
								& e,p DIS							& Photon-Gluon-Fusion ($80-90\%$)	& Internal $\vec{N}$ target				\\
\hline
\bf{Kinematics}	& $1.8< Q^2 <(1.8\times 10^{4})$				&  $5<Q^2_G<100,\sqrt{s_{ij}}>10$ 						&		\\
\bf{Range}	&  $(5.5\times10^{-5})< x < 1$
& 0.0015 - 0.3 
& $0.1<x_G<0.4$ 			\\
\hline
\bf{Theoretical}						& $\Delta G(x,Q^2)~$ \& $\int \Delta G$ 		& LO calculations for ${\cal A}_{\vec{e}\vec{p}}^{e(2~jets)}$ 	& Onset of pQCD 	\\
\bf{Basis}							& from pQCD at NLO	 				& Lack of NLO calculations for				& for $\gamma+(X)$;	\\
\bf{\&}								&							& polarized cross sections				& pQCD for $J/\psi+(X)$	\\
\bf{Uncertainties}						& minimum uncertainties					& and for Monte Carlo					& Should know $\Delta q$ \\
\hline
\bf{Kinematical}						& $y>0.01~\theta_{e'}>3^o$				& $p_T>5$, $\mid \eta \mid<2.8$		& $2\leq p_T\leq 8$	\\
\bf{Constraints}						& $Q^2>1$, $E_{e'}>5$					& $0.3<y<0.8$				& $-1.5 \leq \eta \leq +1.5$	\\
\hline
\bf{Experimental}						& \multicolumn{3}{|c||}{Polarization of $820~$GeV protons in HERA and measurement of proton polarization}			\\
\bf{Problems	}						& 						&					&			\\
								&						& Gluon Compton $2$-jet background							& Identify direct $\gamma's$ \\
\hline
\bf{Statistical Error} & for ${\cal L}=200$ & for ${\cal L}=500$ 
&  for ${\cal L}=240$ \\
\bf{on} ${\cal A}$ & $\delta {\cal A} = 10^{-3}$ to $0.1$ 
& ${\cal A}=$~few$\%$, $\delta {\cal A}<(0.2~to~1\%)$	
&$\delta {\cal A} = 0.003 - 0.05$	\\
\hline
   & Relative error on $\int\Delta G$ &	for ${\cal L}=500$ 
& for ${\cal L}=240$	\\
\bf{on} $\Delta G/G$ & $25(20)\%$ with ${\cal L} = 200(1000)$	
& $\delta(\Delta G/G)$= $0.007-0.1$ & $\delta(\Delta G/G) = 0.03 \div 0.4$	\\
\hline
\bf{Systematics}						& \multicolumn{3}{|c||}{Measurement of $P_e,P_p$ ($\pm5\%$). False asymmetries small since can provide any sign}	\\
								& \multicolumn{3}{|c||}{of $P_p$ for any bunch, and with a spin rotator can change $P_p$ sign of all bunches.}	\\
\hline
\bf{Status}							& \multicolumn{3}{|c||}{Study of polarized protons at HERA; pre-proposal stage}					\\
\hline								 
\bf{Time Scale}							& \multicolumn{3}{|c||}{$\geq~$Year 2003}									\\ 
								& \multicolumn{2}{|c|}{HERA operational with $27~$GeV $\vec{e}$ ; H1 \& ZEUS detectors operational}		&			\\
\hline
\bf{Remarks}							& Low $x$ behaviour of				& $x_G$ is directly measured		& Need (new) HERA-B 		\\
  & $g_1^p(x,Q^2)$ of great interest		& over a wide kinematic range		& type detector 		\\
\hline\hline
\end{tabular}
\hfil
\end{turn}

\end{document}